\documentclass[10pt, conference,letterpaper]{IEEEtran}
\usepackage{amsmath,amsfonts}
\usepackage{algorithmic}
\usepackage{algorithm}
\usepackage{array}
\usepackage{subfig}
\usepackage{textcomp}
\usepackage{stfloats}
\usepackage{url}
\usepackage{verbatim}
\usepackage{graphicx}
\usepackage{cite}
\begin{document}

\date{}

\title{Adaptive Hopping for Bluetooth Backscatter using Commodity Edges\\
}

\author{\IEEEauthorblockN{Maoran Jiang and Wei Gong}
	\IEEEauthorblockA{University of Science and Technology of China\\
		mrjiang@mail.ustc.edu.cn, weigong@ustc.edu.cn\\	
}
}


\maketitle

\begin{abstract}
Channel hopping is essential to BLE backscatter as commodity BLE switches channels frequently during transmission to overcome interferences in busy radio environments. Existing Bluetooth backscatter systems, however, suffer from slow responses to excitation change and poor control of the target channel. To address these issues, this paper presents ChannelDance, a BLE backscatter system that utilizes a low-latency edge server to achieve fast and accurate hopping. Specifically, we show that the backscattered channel relies on the excitation channel and tag toggling frequency. By identifying excitation frequency, the tag can achieve accurate hopping with a dynamically configured clock. Further, we introduce a low-latency architecture, which is centralized, asynchronous, and equipped with high-speed interfaces. This architecture supports the tag to respond to excitation changes fastly. We prototype the ChannelDance tag with FPGA and build the low latency edge server with commodity MCU and off-the-shelf BLE and WiFi radios. Experimental results show that ChannelDance can realize 40 to 40 channel mapping with a median success rate of 93\% and achieve 3.5x goodput gain with channel optimization. Moreover, with adaptive hopping, the ChannelDance tag successfully establishes a connection with commodity BLE.
\end{abstract}

\section{Introduction}

In the past few years, the Internet of Things (IoT) has profoundly affected people's lives, from home, traffic, and entertainment to agriculture, industry, etc. Various entities have been connected to the network, and a tremendous of sensors have been deployed to serve rich applications. For all the efforts we have made, we have a common goal, which is data transmission. Wireless communication has made a great change to our daily life. For example, WiFi has enabled us to access the network with a high throughput link. Bluetooth has made it easy to access these peripherals like keyboards and earphones. However, these radios usually consume lots of power, which requires frequent charging or replacing batteries, which makes them hard to maintain if massively deployed.

To tackle this issue, backscatter technology\cite{liu2013ambient,wang2017fm,hessar2019netscatter,hu2016braidio,vasisht2018body,zhao2018spatial, zhao2018x, wang2012efficient, abari2015caraoke,rostami2018polymorphic,naderiparizi2018towards, varshney2019tunnelscatter,  varshney2020tunnel,liu2020vmscatter, yang2022content} are introduced to greatly reduce the energy consumption of wireless communication and enable a wide range of applications\cite{wang2012efficient, Talla2015PoweringTN, yang2017nicscatter, naderiparizi2018towards, Ma2018EnablingDN, hessar2019netscatter, Jang2019underwater,   Abedi2020WiTAGSW,    Mazaheri2021MmTag}. It leverages the fact that the carrier generating is an energy-hungry part of active radios. Thus instead of generating carriers, backscatter modulates data on ambient carriers to save energy. RFID\cite{buettner2011dewdrop} is one of the mature backscatter systems. A typical RFID system consists of several tags and an RFID reader. The reader generates a carrier wave, and the tag backscatters the carrier back, with data modulating on it. The reader decodes the backscattered signal and decodes the modulated data of the tag. This greatly reduces the power consumption of the tag. However, the dedicated device of the reader is expensive and hinders the development of backscatter technology.

\begin{figure}[t]
	\centerline{\includegraphics[width=0.95\linewidth]{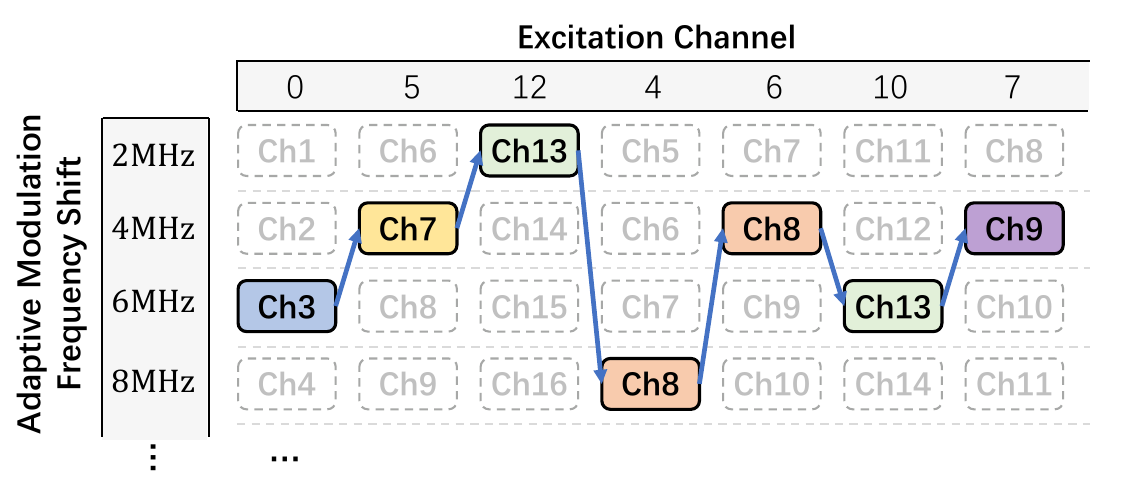}}
	\caption{ChannelDance design. The excitations are on diverse channels and ChannelDance tag performs fast and accurate frequency shift to the excitations to generate target packets on adaptive channels.}
	\label{design}
\end{figure}

Thus lots of efforts have been made on exploring backscatter on commodity radios, for example WiFi\cite{zhang2016hitchhike,bharadia2015backfi,kellogg2014wi,kellogg2016passive, dunna2021syncscatter}, BLE\cite{ensworth2017ble,zhang2017freerider}, ZigBee\cite{zhang2017freerider,iyer2016inter}, etc. HitchHike\cite{zhang2016hitchhike} introduces WiFi backscatter using commodity receivers. FreeRider\cite{zhang2017freerider} enables compatibility with commodity BLE devices. RBLE\cite{zhang2020reliable} enables direct interaction with commodity BLE. And IBLE\cite{zhang2021commodity} improves the modulation quality BLE backscatter. These works aim to enhance various characteristics, like high throughput, long distance, and good compatibility. Since BLE adopts a simple modulation technique and is widely available around us, a majority of the work is exploring compatibility with commodity BLE devices and trying to turn BLE backscatter into general-purpose communication.
The state-of-the-art BLE backscatter systems \cite{ensworth2017ble,zhang2017freerider, iyer2016inter, zhang2020reliable, zhang2021commodity} have utilized BLE excitation packets as single tones and regenerated new packets on the target channel. This greatly improves compatibility but there is still a gap between BLE backscatter and commodity BLE. BLE adopts channel hopping to overcome interference in busy radio environments. This characteristic has been rarely considered. 

On the one hand, state-of-the-art BLE backscatter systems rarely consider excitations hopping during backscattering. FreeRider\cite{zhang2017freerider} adopts the codeword translation technique, which modulates data on BLE excitation packets and decodes out tag data by comparing the backscattered data and excitation data. It did not consider the hopping excitations. RBLE uses the single-tone parts of excitation packets as carriers, it cannot handle excitations on different channels too.

On the other hand, these BLE backscatter systems do not have designs for target channel hopping. FreeRider modulates data on the payload. While the access address and header field of backscattered packets are the same as the excitation packets. However, the commodity device adopts whitening in transmitting and receiving, so the receiver, which is running a BLE stack, cannot receive packets on different target channels, because the whitening seed is related to the channel index. Thus FreeRider cannot achieve channel hopping on target channels. RBLE supports backscattering on different target channels, but it does not have a design of hopping logic.

To address the above issues, we present ChannelDance, a BLE backscatter system that utilizes a low-latency edge server to achieve fast and accurate hopping among different excitation packets. As shown in Fig. \ref{design}, the different excitations are on diverse channels. The tag applies adaptive frequencies shift to the excitation packets, generating fast and accurate hopping packets on different channels. To achieve ChannelDance in practice, we need to address two main challenges.

\begin{itemize}
	\item [1)]\emph{How to achieve accurate hopping with different excitations?}\\
	The backscattered channel is determined by both the excitation frequency and modulation frequency. Compared to the modulation frequency, excitation channels are less controllable. The excitor can take different channel selection algorithms, and meanwhile, there are also various algorithm settings, which are hard for the tag to predict. Moreover, transmission quality may vary among different channels, and tag needs to identify these channels with good qualities. To address this, we adopt an edge server to forward the channel information to the tag and the ChannelDance tag adapts the modulation clock on demands.
	\item [2)]\emph{How to achieve fast hopping?}\\
	The commodity BLE devices can switch channels at a high rate, which can be more than 100 times in one second. The edge server needs to update the excitation information at such speed; otherwise, the packets could be backscattered wrongly. RBLE proposes Packet Length Modulation (PLM) to convey downlink data. However, PLM suffers from a high delay of time of seconds, which can not adapt to the high hopping rate. ChannelDance proposes a low-latency architecture that can forward information to the tag at high speed.
\end{itemize}

We prototype ChannelDance tag using FPGA and the edge server with off-the-shelf MCU and radios. We conduct extensive empirical experiments, and the results show that

\begin{itemize}
	\item The maximum success rate for channel hopping of ChannelDance is 100\%. Moreover, it dynamically achieves 40 to 40 channel hopping with a median success rate of more than 93\%.
	\item ChannelDance can actively scan the total 40 channels and identify target channels with PER less than the median value and achieves 3.5x goodput gain over the bottom 20 percent with channel optimization.  
	\item ChannelDance edge server can forward the channel information to tag with a minimum delay of 5.8 $ms$, 400x lower than RBLE hopping delay. Moreover, ChannelDance accurately achieves BLE channel selection algorithms \#1 and \#2 with this low latency architecture.
	\item With this adaptive hopping design, ChannelDance successfully establishes a connection with commodity BLE devices.
\end{itemize}

\begin{figure}[t]
	\centerline{\includegraphics[width=0.9\linewidth]{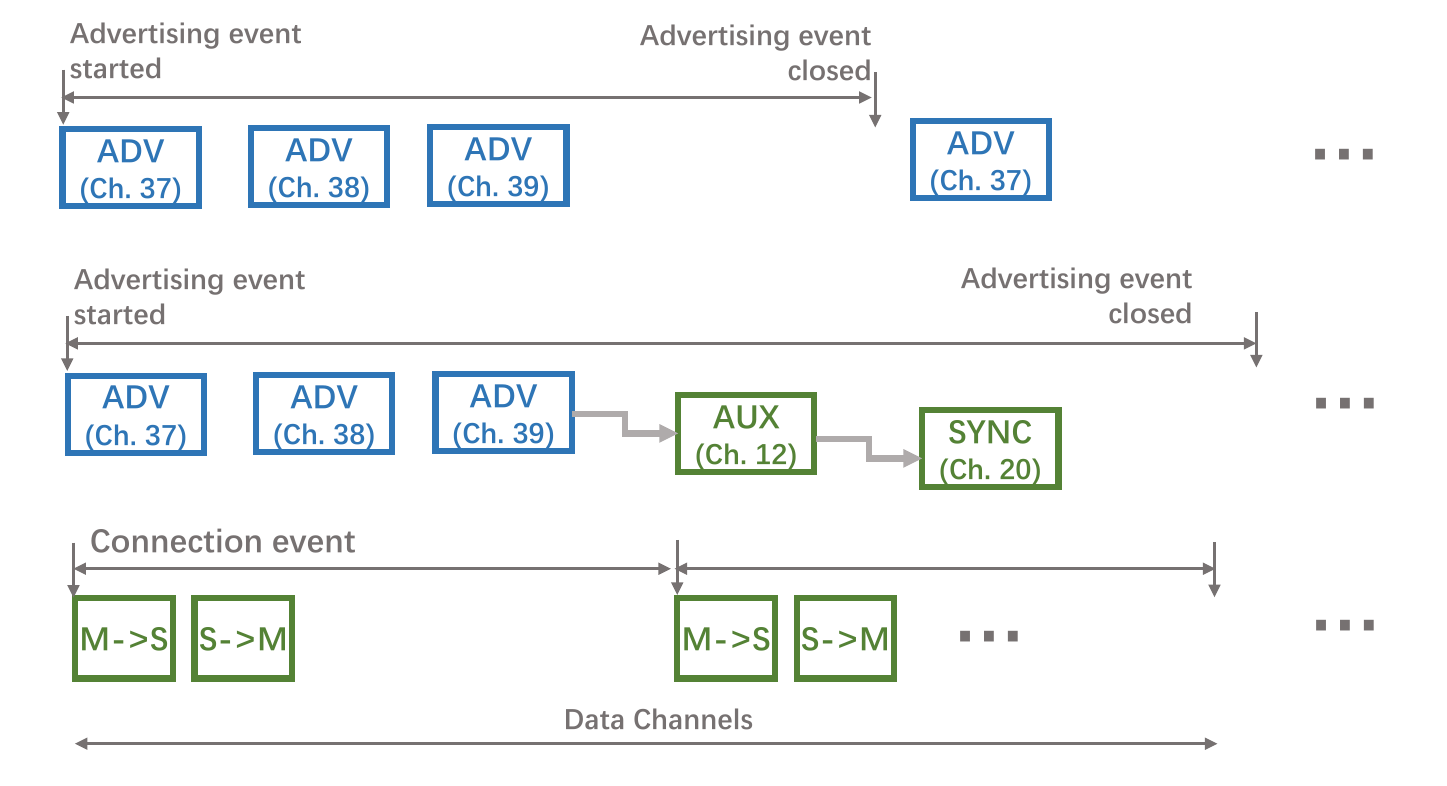}}
	\caption{Channel hopping is adopted in BLE advertisement, extended advertisement and connection.}
	\label{primer}
\end{figure}

\section{Primer Knowledge}
\subsection{Bluetooth Low Energy}
Channel hopping is one of the key features of Bluetooth Low Energy. Like other communication techniques in the ISM band, e.g., WiFi, ZigBee, BLE divides the frequency band into 40 channels for transmission to avoid interferences. Meanwhile, BLE also introduces channel hopping to tackle interferences. We introduce the feature from two perspectives: advertisement and connection.

In BLE 4, three channels are used for advertising. BLE devices advertise at advertising events. In each event, devices broadcast the same data on the three advertising channels one by one. And other BLE devices scan on the three channels. BLE 5 introduces extended advertisement to enhance advertising. Besides the three traditional advertisement channels, BLE 5 also performs secondary advertising on the 37 data channels. The device first advertises on the traditional advertisement channels and then the secondary channels, which ensures compatibility with BLE 4 devices.

BLE connection involves channel hopping in the whole process. In each connection event, the pair of devices send data to each other at precisely timed intervals. At the start of each connection event, the devices do channel hopping, which selects a channel from the available data channels based on channel selection algorithms. So during connection, the transmission will be carried over a series of changing channels. Meanwhile, the involved devices also maintain a channel map to mark the channels as used or unused based on the channel quality. With negotiation during connection, those bad channels could be excluded.

\subsection{Backscatter technology}
In active communication, the device generates radio waves to transmit data, which consumes lots of energy. To achieve an ultra-low power transmission, backscatter technique reflects the ambient signal instead of generating the signal itself. During reflection, backscatter nodes will change the frequency, phase, and amplitude of the radio signal to modulate data. To enable compatibility with commercial devices, a common method used in state-of-the-art backscatter systems is called codeword translation. Codewords are signal symbols that denote different data. 

\section{ChannelDance Design}
We first give an overview of ChannelDance framework, then introduce how we enable adaptive hopping, design the low-latency edge server, and modulate tag data.

\subsection{Overview}
\begin{figure}[t]
	\centerline{\includegraphics[width=0.95\linewidth]{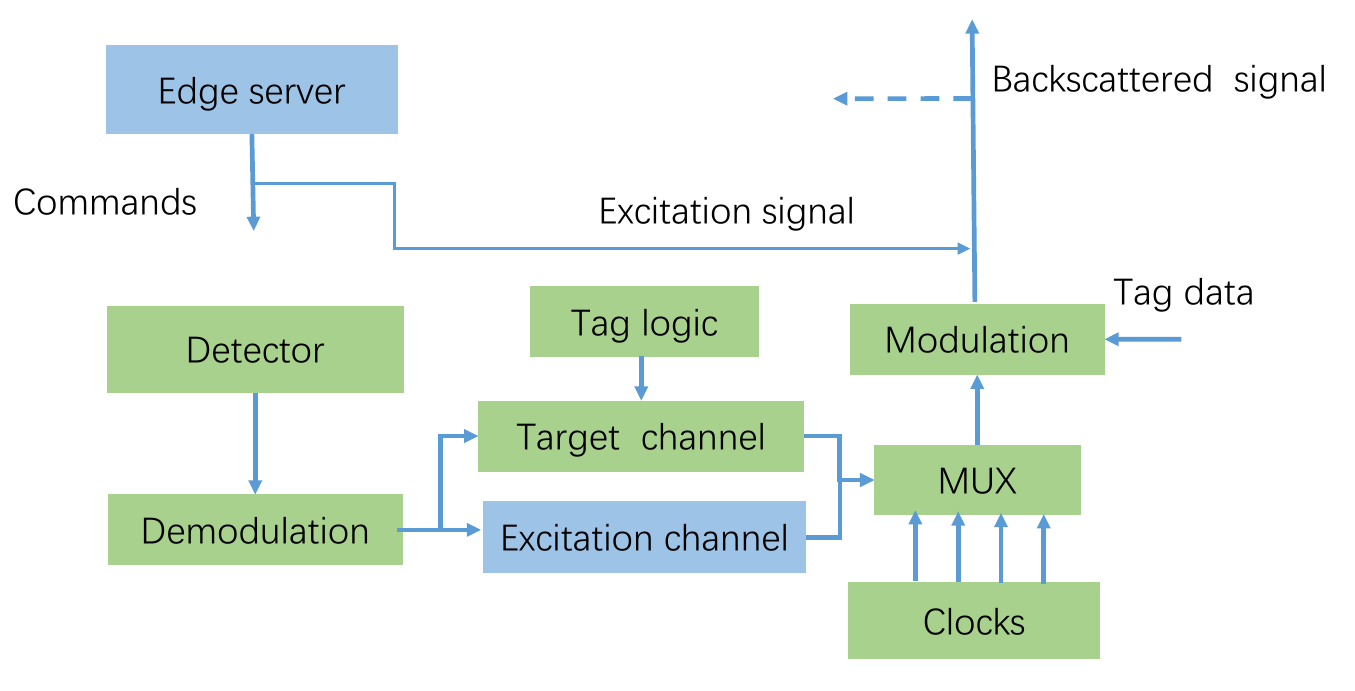}}
	\caption{ChannelDance framework. After receiving the data from the edge server, the ChannelDance tag extracts the excitation channel information, and choose the proper clock to do modulation. Edge server can also evaluate the channel quality and update the channel map to the tag.}
	\label{overview}
\end{figure}

As shown in Fig. \ref{overview}, the ChannelDance system mainly consists of two parts: edge server and tag. The edge server provides excitation signals on different channels. Meanwhile, it also supports sending commands to the tag, which can be excitation channel information or ambient BLE data. Tag decodes the downlink data and gets the frequency of the excitation channel. Then it calculates the required frequency shift based on the excitation and target channel and configures the clock. ChannelDance tag can achieve fast and accurate channel hopping. Meanwhile, the edge server can also evaluate the performance of different backscatter channels and enable the tag to adaptively disable unwell channels.

\subsection{Accurate hopping}
Channel hopping is one of the key features of BLE. BLE adopted it to tackle the interference since 2.4 G bands are also widely used by other communication devices, like WiFi, and ZigBee. BLE adopts channel selection algorithms to determine the target channel during transmission.  To enable interactions with commodity BLE devices, the tag needs to support adaptive hopping. Meanwhile, BLE also evaluates the channel quality, and it can avoid transmission on the channels that suffer from low quality. BLE backscatter faces the same interferences as commodity BLE. It is necessary for the tag to know the channel quality and transmit on reliable channels.  A BLE backscatter tag that supports adaptive hopping should enable the following operations.
\begin{itemize}
	\item Hopping. A tag can backscatter excitation packets to different channels. It can also hop along the channel sequences generated by the BLE channel selection algorithms.
	\item Optimization. A tag can sense the quality of different channels and maintains a used map. Those channels with poor quality can be excluded during transmission.
\end{itemize}

The state-of-the-art BLE backscatter systems, FreeRider\cite{zhang2017freerider}, RBLE\cite{zhang2020reliable}, and IBLE\cite{zhang2021commodity}, fail to support the above two operations.
RBLE is not able to sense the carrier frequencies and cannot adapt to excitation packets from different channels. FreeRider and IBLE do not provide a hopping design. 
To achieve adaptive hopping, first, the tag needs to know the carrier frequency. BLE backscatter shifts the excitation signal to the target channel by operating the RF switch using a clock. The frequency shift depends on the frequency of excitation packets and the target channel. Second, the tag needs to implement the channel selection algorithm logic. And the evaluation of channel quality should also be maintained. To support these, a tag downlink is required.
Since the backscatter tag cannot directly decode frequency-modulated signals, state-of-the-art backscatter systems \cite{rostami2018polymorphic, li2022passive, guo2022saiyan} utilize the envelope decoding. 
To detect the excitation channel, a possible solution is to use a Surface Acoustic Wave (SAW) filter\cite{guo2022saiyan} to sense the frequency. For 2.4GHz signal, we find QPQ1905 IoT bandBoost Filter. However, it has a bandwidth of 25MHz, which is far larger than the 2MHz of BLE channels. It is hard to detect the channel of excitation signals. 
To acquire excitation frequency and channel quality, we introduce an edge architecture, consisting of an ASK downlink for tag, which will be detailed in Section \ref{section:edge}.

\begin{figure*}[t]
	\centerline{\includegraphics[width=0.99\linewidth]{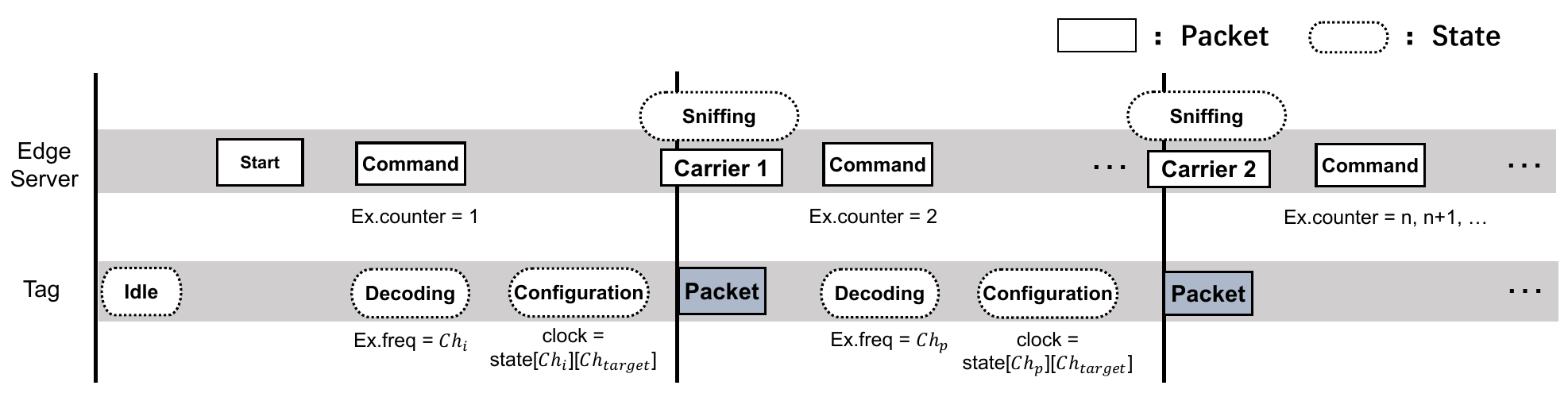}}
	\caption{The edge server sends a start signal to notify the tag that it will start to generate excitations. Before each excitation packet, the edge server will send the channel information to the tag. The ChannelDance tag decodes it and set the modulation clock based on hopping sequence. Meanwhile, the edge server can receive the backscattered packets to assess the channel quality and update the used channel map of the tag if necessary.}
	\label{accurateHopping}
\end{figure*}

 In ChannelDance, the edge server provides the excitation signal and channel frequencies for the tag. There can be two cases. In the first case, the excitation packets stay in one channel, e.g., 37. The edge server will send the channel index to the tag before the start of excitations and update the channel information if the excitation channel changes. In another case, the excitation packets are hopping according to the BLE specification, and the channel is determined by the packet index. As shown in Fig, \ref{accurateHopping}, the edge server gets the packet counter according to the selection algorithm and sends it to the tag before excitaion. 
When tag receives the downlink packet, it checks the correctness. If there are no errors, the tag updates the excitation packet counter. If the downlink command is missed or fails the check, tag increments the counter itself. With this packet counter, tag calculates the excitation frequency through the channel selection algorithm.

To backscatter the excitation packets to different target channels, the tag changes the clocks with different frequencies to control the RF switch. BLE has 40 channels, and to achieve a 40 to 40 channel mapping, we need 78 sets of frequencies. By utilizing the double sideband modulation, we can reduce the frequency number to 39. However, it is impossible to generate 39 sets of frequencies at one time. We adopt dynamic channel reconfiguration to generate a set of frequencies at a time. It leverages the fact that we can modify the frequency output by changing the values of clock registers. 
\begin{figure}[t]
	\centerline{\includegraphics[width=0.8\linewidth]{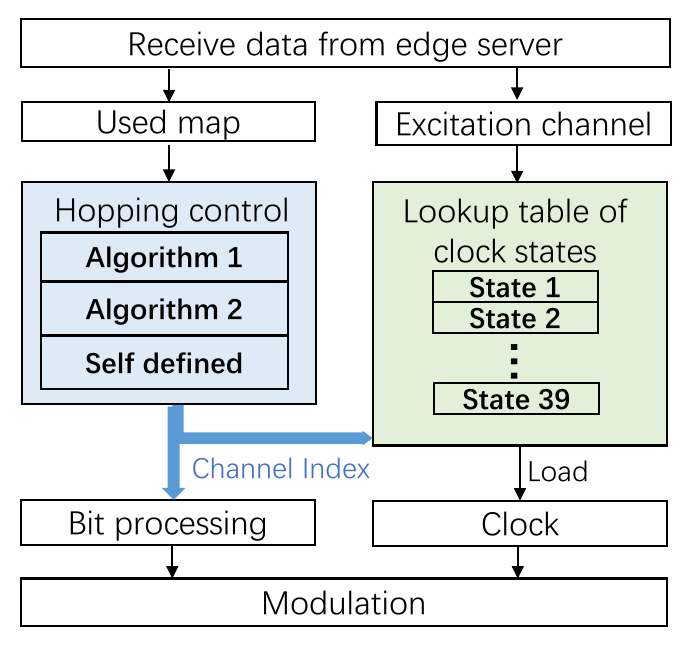}}
	\caption{The logic of hopping control in ChannelDance tag selects the target channel, which is used to determine the clock states with excitation channel index. The registers of the selected clock state will be loaded to generate a modulation clock. Meanwhile, the target channel index is also used for PHY bit processing, e.g., whitening.}
	\label{tagLogic}
\end{figure}
The output clock is generated as shown in the following equation:
$$CLK0 = (CLK\_INPUT * MUL / DIV) / CLK0\_DIVIDE$$
where $CLK\_INPUT$ is the reference clock, $MUL$ and $DIV$ are the multiplication and division factors for the reference clock respectively. $CLK0\_DIVIDE$ is an extra division factor only for output $CLK0$.
The registers corresponding to these factors are precomputed during compilation. We group the clock registers needed in shifting one pair of channels as states. And we use the IP core in Vivado to generate the factors for our 39 sets. 
With 39 states stored, we use a lookup table $state[excitation][target]$ to hit the state that hopping needs as shown in Fig. \ref{tagLogic}.

These clock registers control the PLL circuit to generate the target clock. They also consume resources in FPGA, which should be considered. According to \cite{xapp888}, a configuration state consumes about $9 + 2n$ words and each word consists of 39 bits, where $n (< 7)$ denotes the num of clocks in one state. When specifying 4 clocks at a state, we need to store 663 bits. ZYNQ takes one 6-LUT to store 64 bits. Thus one state consumes 10 LUTs on average, far less than the 17600 LUTs in ZYNQ.

With dynamic clocks, the tag can backscatter excitation packets to 40 channels. While the channel quality could vary from different channels, we enable the tag to scan the whole channel and exclude the channels with poor quality. Thus we enable the edge server to receive the backscattered packets and evaluate each channel’s performance. After the scanning, the edge server will update the map to the tag, which indicates channels with good performance.
Being able to transmit on 40 channels, it is important to tag to manage the hopping sequences too. BLE adopts two channel selection algorithms. The device switches the channel frequency at the start of each event. In algorithm \#1, the channel is determined by the used channels and hop interval. The channel is usually sequentially selected. The hop interval and map are set by tag on demand. For algorithm \# 2, the channel is determined by the used channels, and access address. The channel sequence is usually semi-random. The access address is also set by the tag and algorithm \# 2 can be achieved in the tag too.

\subsection{Low latency edge server}\label{section:edge}

In the above, we introduce the edge server to assist the tag to get the excitation frequency and enable channel hopping. Besides conveying information to enable accurate hopping, the edge server also needs to forward data fast. BLE backscatter leverages ambient BLE packets as excitations. And the excitation packets can hop at a high rate. In BLE 4, the advertisement packet can hop at the three advertising channels with an interval of less than 10 $ms$. In BLE 5, the newly introduced extended advertisement can achieve a hopping rate of 7.5 $ms$. And in BLE connection, channel hops at the start of each connection event, which has a minimum interval of 7.5 $ms$ too. These high-rate hoppings require the edge server to enable low latency forwarding.

The edge server receives the packet data from commodity device, processes it, and then sends it to the tag via downlink. Let’s first analyze the cause of the delay during the edge server forwarding data.  The first part is the receiving delay of the BLE module. The BLE module in the edge server needs to receive data and then send it to the controller, which takes time. The second part of the delay is the time required for the controller to process the data and send it to other modules. The third part is the ASK signal-generating part. The transmission module needs to wake up the radio and send out an ASK signal, which takes time too. The last part is the decoding time taken by the tag. We introduce a low-latency edge server architecture based on the above analysis.

\begin{figure}[t]
	\centerline{\includegraphics[width=0.98\linewidth]{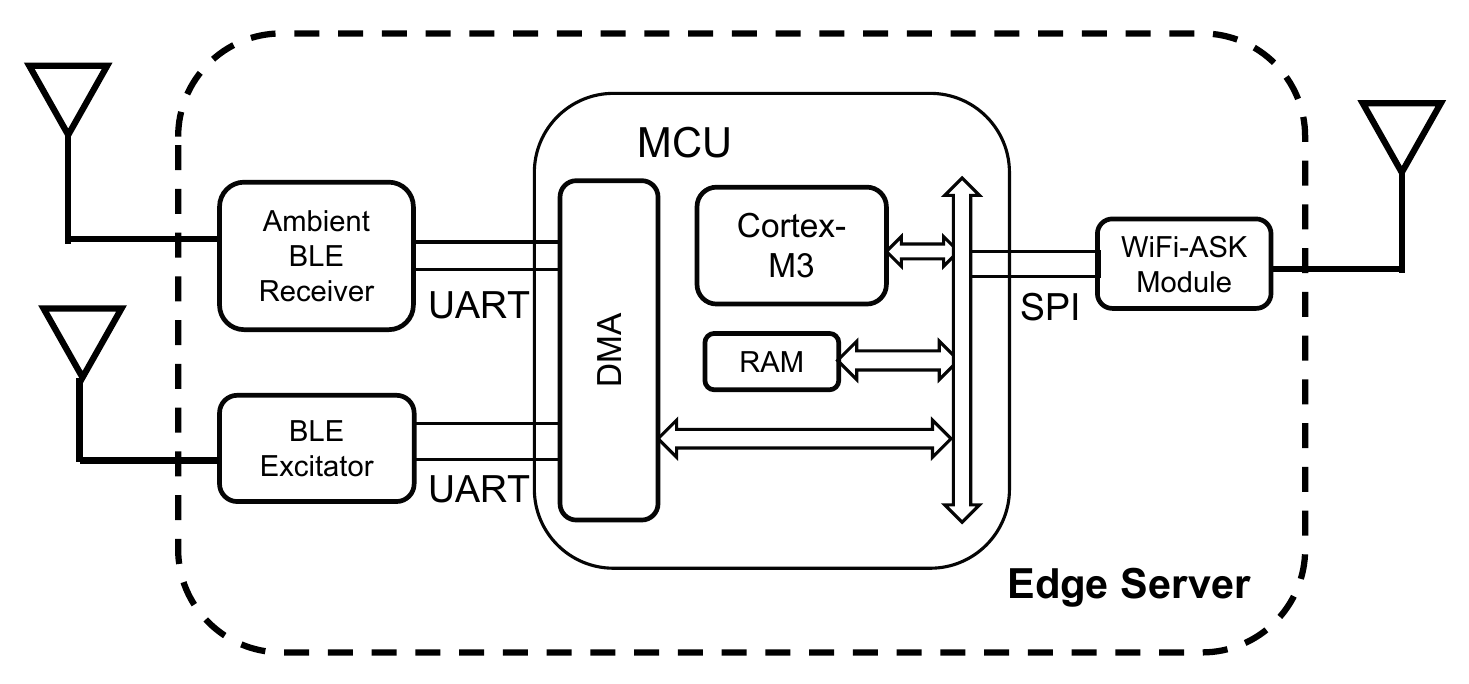}}
	\caption{ChannelDance edge server utilizes a centralized and synchronous architecture using off-the-shelf radios and MCU.}
	\label{edgeArc}
\end{figure}

The edge server consists of different modules to undertake different functions as shown in Fig. \ref{edgeArc}.
\begin{itemize}
	\item Ambient BLE receiver. This module is used to sniff the ambient BLE packets in the environment. It can extract the data and channel information.
	\item Excitor. This module is used to provide the tag periodic and reliable excitation packets. It outputs the packet information to the controller before the generation of each excitation packet.
	\item ASK generator. It receives the commands and data from the controller in the edge server and sends downlink data to the tag. We use a WiFi radio to generate the ASK-like signal by using the modulation scheme in Interscatter \cite{iyer2016inter}. 
	\item Controller. The controller is the brain of the edge server. It forwards data between different modules and coordinates them to work in order. 
\end{itemize}

 The ambient BLE receiver of edge runs a micro BLE stack, which is small and flexible, to sniff the ambient BLE packets. Meanwhile, it uses a universal asynchronous receiver /transmitter (UART) to transmit data to the controller of the edge server. UART is commonly used for serial communication over BLE devices. Its data rate is configurable, and 115200 is the most common rate. Different devices may support different higher rates. 115200 baud rate would be slow for our edge server. If the BLE has tens of bytes to transmit, this low baud rate will consume milliseconds of time. Thus we take a higher baud rate of 1000000, which would reduce the delay to hundreds of microseconds. For the excitation module, we adopt a UART of 1000000 baud rate for the transmission of excitation information between modules, too. For the ASK module, the aggregated data from different modules would be transmitted all through it. We use SPI for the communication between the controller and the ASK module. SPI is also a common communication interface that supports high data rates. We set the speed to 4 MHz, reducing the common transmission time to tens of microseconds.

ChannelDance's edge adopts a common architecture that is used by commercial backscatter systems (such as Wiliot and Jeeva Wireless). This architecture comprises an edge device and multiple tags.
And importantly, the edge device in our system incorporates off-the-shelf radios (BLE and WiFi). It transmits signals on BLE and WiFi channels following specifications, without occupying an extra spectrum and causing interference. 
To further reduce the channel occupancy, the edge can share the signal information with tag in a low-duty cycle. This is because our edge sends the metadata of carrier signal instead of raw frequency information. The tag implements the BLE channel selection algorithm uses simple logic and it can reliably calculate and predict future signal information using the metadata within a period. With this design, the edge transmits data in a low-duty cycle.


\subsection{Tag uplink}
ChannelDance regenerates the BLE packets with the data from the tag and the edge server. Then it adds 3 bytes CRC to the end of the packet and does whitening with the target seed. Thus we get the raw packet data ready for modulation. 

ChannelDance adopts phase modulation, which is introduced by IBLE \cite{zhang2021commodity}, on single tones. IBLE proposes that the integral of frequency over time leads to the change of phase. So IBLE regenerates packets on a continuous wave using phase modulation. It shows that a phased increase of 90 degrees equals the integral of $+250 kHz$ over time. ChannelDance applies it to the single-tone packets. We have
\begin{equation}
	\begin{split}
		r(t) &= sin(2\pi f_0 t)\times S_{tag}(f_ot, \Delta \theta) \\
		&= \frac{4}{\pi}\sum_{n=1,3,5...odd}^{\infty}\frac{1}{n} sin(2\pi f_0 t)sin(2\pi nf_tt + \Delta \theta) \\
		&= \frac{2}{\pi}(cos(2\pi(f_0+f_t)t + \Delta \theta)-cos(2\pi (f_0-f_t)t - \Delta \theta)) \\
		&= \frac{2}{\pi}(cos(2\pi f_{R0}t + \Delta \theta)-cos(2\pi f_{L0}t - \Delta \theta))
	\end{split}
\end{equation}
where $f_0$ is the single tone on excitation channel,  $f_t+\Delta \theta$ is the frequency shift tag toggles. IBLE uses square wave $S_{tag}$, which can be presented using Fourier series and we eliminate the high order harmonics. And due to double side modulation, we got two backscattered signals $cos(2\pi f_{R0}t + \Delta \theta),cos(2\pi f_{L0}t - \Delta \theta)$, where $f_{R0}, f_{L0}$ denote the frequency of bit 0 on right band and left band respectively. When backscattering BLE symbol 0 on both bands, we apply $\Delta \theta = 0$ to the RF switch. When backscattering BLE symbol 1 on both bands, we apply $\Delta \theta = \pi, -\pi$ respectively, indicating a frequency deviation of $500 kHz$. For example, with excitation on channel 4, we can generate expected BLE symbols on channels 6 and 2 with the 4MHz clock, and the clock phase of RF switch in each symbol is: 
$$tag\enspace bits: 0, 1, 1, 0,1, 0,1, 1$$
$$channel\enspace6:0, \pi, 2\pi, 2\pi, 3\pi, 3\pi, 4\pi, 5\pi$$
$$channel\enspace4:0, -\pi, -2\pi, -2\pi, -3\pi, -3\pi, -4\pi, -5\pi$$
$$rf\enspace switch:0, \pi, 0, 0, \pi, \pi, 0, \pi$$

Since $-\pi$ equals $\pi$, both channels get the same control sequences. With the above modulation technique, ChannelDance regenerates packets on different target channels.

\section{Implementation}
We build a prototype of ChannelDance using commodity radios, MCUs, and FPGAs. Our off-the-shelf prototype consists of a tag and an edge server. The details are as follows.



\textbf{Tag prototype}. The ChannelDance tag mainly consists of two parts: RF front-end and FPGA processing unit. The RF front-end consists of an envelope detector and an RF switch. We use AD9235 to sample the output of the envelope detector for further processing in FPGA. An ADG902 RF switch is connected to FPGA to do modulation. For the FPGA part, we adopt a ZYNQ XCZ7010 for function verification, which is used for signal processing, tag logic, and RF switch control.  

\textbf{Edge server.} The ChannelDance edge server mainly consists of four parts: excitation generator, commodity BLE receiver, WiFi-Ask transmitter, and MCU. The excitation generator is built on NRF52840, running Zephyr periodic advertisement application. We use TI CC3200 to transmit the WiFi-ASK signal. CC2650 is a commodity receiver to receive ambient BLE packets and evaluate tag modulation performance. We use NUCLEO-F411RE MCU, which has sufficient UARTs and SPIs, to build the control unit of the edge server. The edge will have a low cost in mass production due to the use of off-the-shelf chips.

\textbf{Ambient BLE device.} We use CC2540 as an ambient BLE device. Different from the BLE module operating on the self-defined mode in the edge server, this ambient BLE device operates a commodity BLE stack. ChannelDance tag can build a BLE connection with this commodity device and send tag data.

\textbf{Experiment setup.} The experiments are all conducted in an indoor environment. In experiments on hopping success rate and packet error rate, we use CC1352 as an excitation generator instead of NRF52840, because the TI devices have test modes and it is more efficient to be configured to conduct a comprehensive mapping evaluation. 
NRF52840, running the BLE5 stack, is used for the hopping algorithm experiment and connection application.

\begin{figure}
	\centering
	\begin{minipage}{0.47\linewidth}
		\subfloat[{Tag prototype.}]{\includegraphics[width=\linewidth]{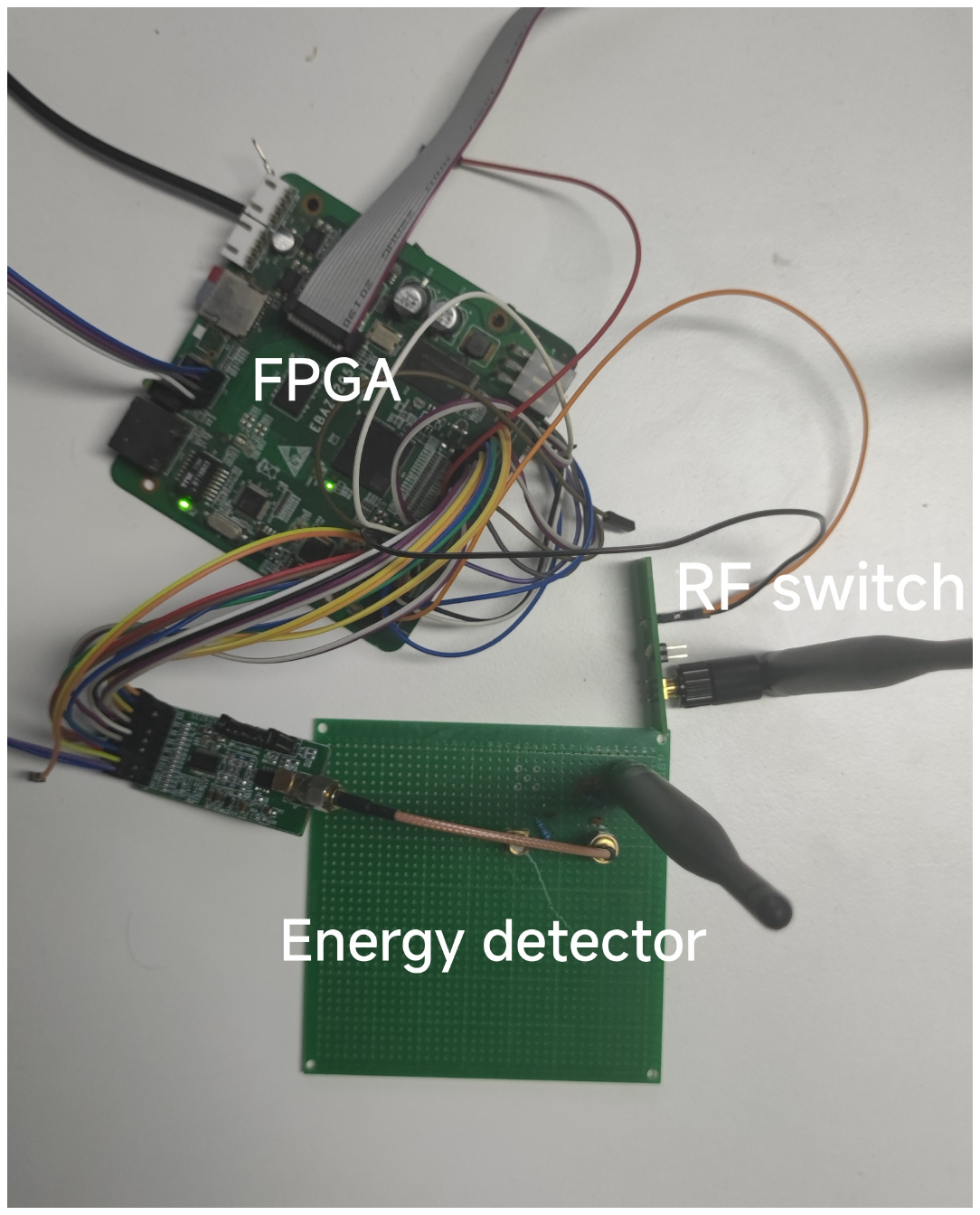}
			\label{tagProto}
		}
	\end{minipage}
	\begin{minipage}{0.465\linewidth}
		\subfloat[{Edge server prototype.}]{\includegraphics[width=\linewidth]{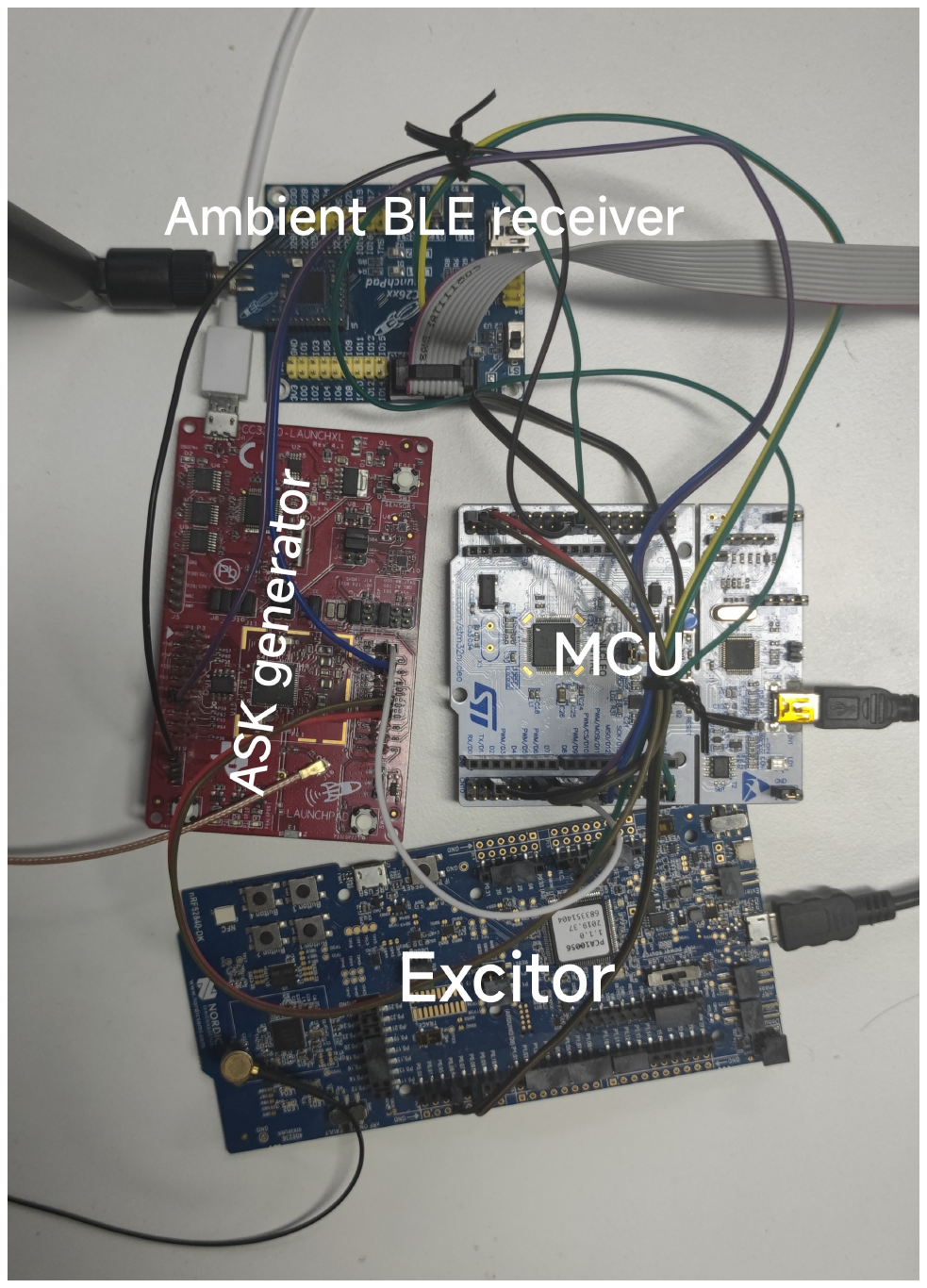}
			\label{edgeProto}
		}
	\end{minipage}
	\caption{ChannelDance prototype.}
	\label{photo}
\end{figure}

\textbf{Metric.} One of the performance metrics is the mapping success rate. The excitation generator generates a fixed number of excitation packets on different channels, and ChannelDance tag backscatters packets on them with channel hopping. We calculate the reception ratio of the BLE receiver as the success rate. Another metric, goodput, is the data size transmitted during a time interval, which is related to the excitation rate. We fix the excitation rate to evaluate the relative improvement of goodput with our design.

\begin{figure*}[t]
	\centerline{\includegraphics[width=0.97\linewidth]{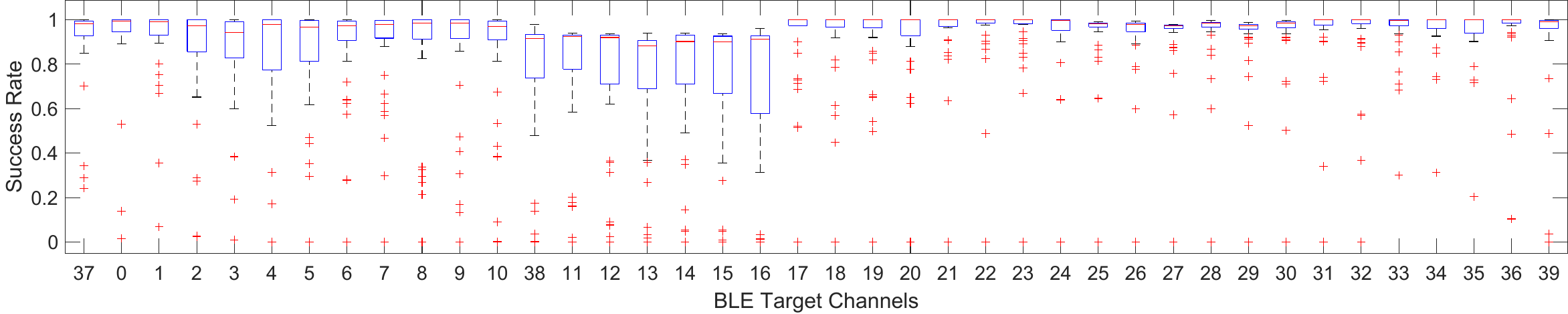}}
	\caption{Success rate distribution of N to 1 hopping in different target channels.}
	\label{NToOne}
\end{figure*}

\begin{figure*} [t]
		\centering
            \begin{minipage}[t]{0.22\linewidth}
			\centering
			\centerline{\includegraphics[width=1\linewidth]{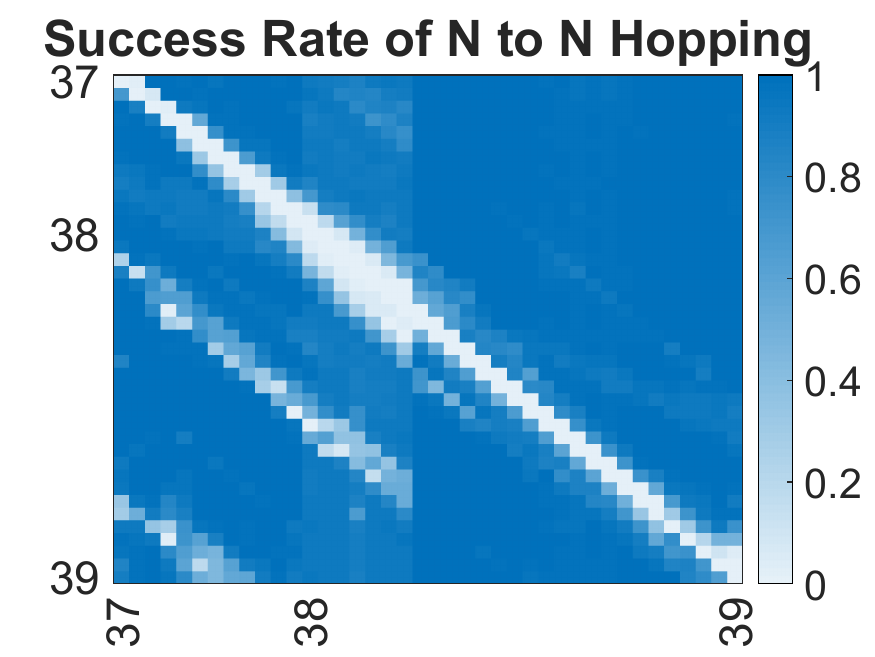}}
			\caption{Traffic pattern of BLE in different locations.}
			\label{heatMap}
		\end{minipage}
   \hspace{0.1in}
		\begin{minipage}[t]{0.32\linewidth}
			\centering
			\centerline{\includegraphics[width=1\linewidth]{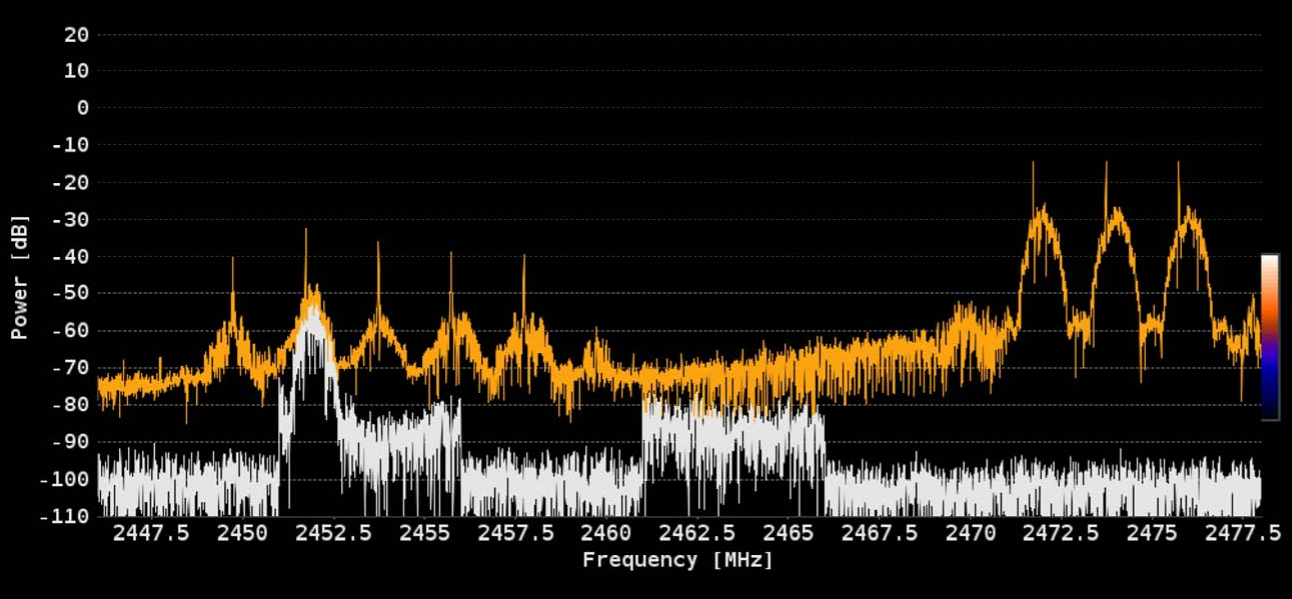}}
			\caption{Traffic pattern of BLE in different locations.}
			\label{spectrumMapping}
		\end{minipage}
  \hspace{0.1in}
		\begin{minipage}[t]{0.3\linewidth}
			\centering
			\centerline{\includegraphics[width=1\linewidth]{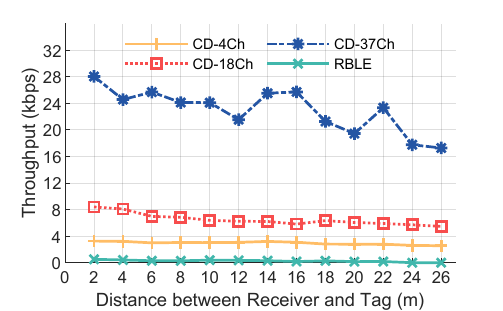}}
			\caption{ChannelDance's and RBLE's throughput under diverse excitations.}
			\label{throughput}
		\end{minipage}
	\end{figure*}

\section{Evaluation}
In this section, we fully evaluate ChannelDance’s performance. We first measure the hopping performance under comprehensive cases and then conduct adaptive hopping experiments.
\subsection{Hopping performance}
To evaluate the hopping performance, we measure success rates, hopping delay, PERs, and goodputs of channel hopping.

\textbf{N to 1 hopping.} We conduct N to 1 hopping for each channel. The excitation generator CC1352 sends 500 packets on each excitation channel; edge server notifies the tag of the excitation channel frequency and target channel. ChannelDance tag backscatters 500 packets from one excitation channel to one target channel and then backscatters packets of the next excitation channel. We conduct this measure for different target channel cases.

Fig. \ref{NToOne} shows the distribution of the success rate on different target channels. Taking Channel 37 as an example, the first quartile (Q1) is 0.9275, which means the success rate of the 75 percentile of excitation channels to this is larger than 92.75\%. The third quartile (Q3) is 0.994, implies that the success rate of the 25 percentile of the excitation channels to this is 99.4\%. 
For all target channels, the Q1 are larger than 0.5795, the medians are larger than 0.882, and the Q3 are larger than 0.9075. 
This implies that 50 percent of ChannelDance hopping for each target channel can achieve a success rate of more than 88.2\%, which means most hopping to the current channel are effective.  
And 75 percent of hopping can achieve a success rate of at least 57.95\%.

As shown in Fig. \ref{NToOne}, there are some outlier values. For most excitation channels, we see a success rate of 0, which is usually on the neighbor channel with a hopping step of 2MHz. This may be related to two reasons. First, the small clock of 2MHz generated directly by MMCM is not accurate enough. The second reason may be that due to the simple architecture of the backscatter tag, the backscattered signal is interfered by the neighbor channel to some extent.



Fig. \ref{NToOne} also shows that the overall success rate of channels 0-10 and 37, 38 are worse than the remaining target channels. It may be the reason that the experiment on the second half of target channels is conducted one day before the first half or the second half channels has better quality. The hopping performance may be affected by the different channel quality. ChannelDance can adaptively choose different excitation channels to improve performance.

\begin{figure*}[t]
	\centerline{\includegraphics[width=0.97\linewidth]{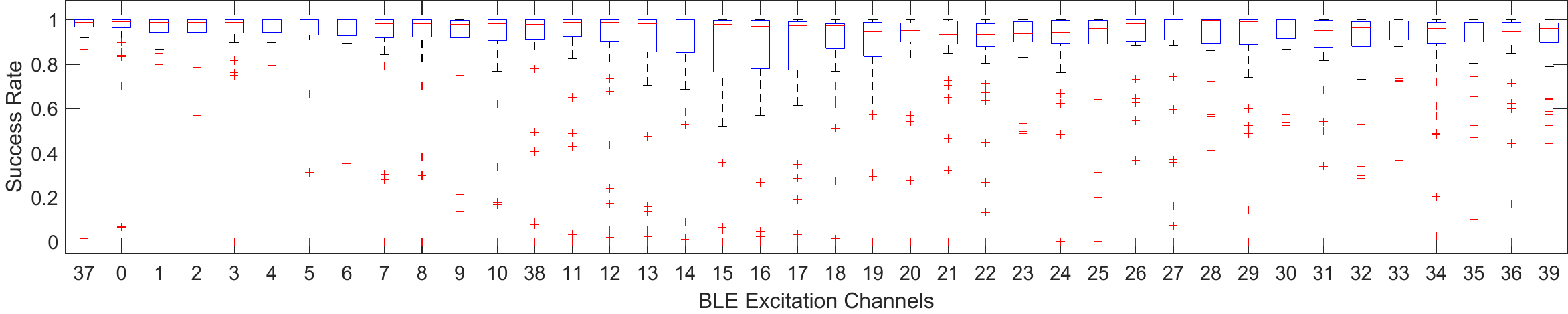}}
	\caption{Success rate distribution of 1 to N hopping in different excitation channels.}
	\label{oneToN}
\end{figure*}

\textbf{1 to N hopping.} We also display the distribution from the perspective of 1 to N hopping. Taking Channel 37 as an example, the first quartile (Q1) is 0.968, which means the success rate of the 75 percentile of target channels is larger than 96.8\%.
Fig. \ref{oneToN} shows that the Q1 are all larger than 0.767, the medians are all larger than 0.934, and the Q3 are all larger than 0.974. This implies that for each excitation channel, 50 percent of  ChannelDance hopping can achieve a success rate of more than 93.4\%. And 75 percent of hopping can achieve a success rate of at least 76.7\%, which means ChannelDance can enable reliable hopping to most target channels.

For each excitation channel,  there are on average about 3 of the channels ChannelDance maps to have a success rate of less than 50\%. This could be caused by some inaccurate dynamic clock on the tag, or by interference. Although some hoppings are having a bad success rate performance, we can exclude this small part of target channels from the used maps to improve the overall performance.

\textbf{N to N hopping.}We show the overall performance of hopping in Fig. \ref{heatMap}. The x-axis denotes the target channel 37, 0, 1, …, 39 from left to right, in the order of frequency, and the y-axis denotes the excitation channel in the same order from top to bottom. The darker color represents a higher success rate. We observe that there is a light line in the diagonal due to hopping failure in the neighbor channel. Meanwhile, there is also a light line in the lower left and upper right. These low success rates are lying in the hopping cases with a clock frequency of around 28 MHz. The bad performance may be caused by the inaccurate clock.

\textbf{Hopping spectrum.} ChannelDance can sense the excitation frequency and adaptively hop to different channels. We observe the spectrum of 3 channels hopping to 5 target channels. The edge server generates periodic single tones on channels 33, 34, and 35. We can see three higher peaks, which are excitation signals, at the right band in Fig. \ref{heatMap}. With edge server assisted, the tag adopts a dynamic clock and adaptively hops on channels 22-26. We can see four lower peaks, which are backscattered signals, at the left band of the spectrum.

\textbf{Throughput comparison.} We also examine the throughput superiority of ChannelDance over RBLE when excitations are from 37 channels. As shown in Fig. \ref{throughput}, ChannelDance utilizes a different number of excitation channels and backscatters them to the data channel. Since CD-37Ch has a utilization of 100\%, it achieves a maximum throughput of 30.69 kbps, which is 53.11x better than RBLE at 1 m of tag-rx distance. When the tag-rx distance is 22 m, the throughput of CD-37Ch is 23.35 kbps, which is 3.9x, 8.3x, and 135.7x better than CD-18Ch, CD-4Ch, and RBLE respectively.

\begin{figure}[t]
	\centering
	\subfloat[Comparison of delay constitution of different edge server architectures.]{\includegraphics[width=0.46\linewidth]{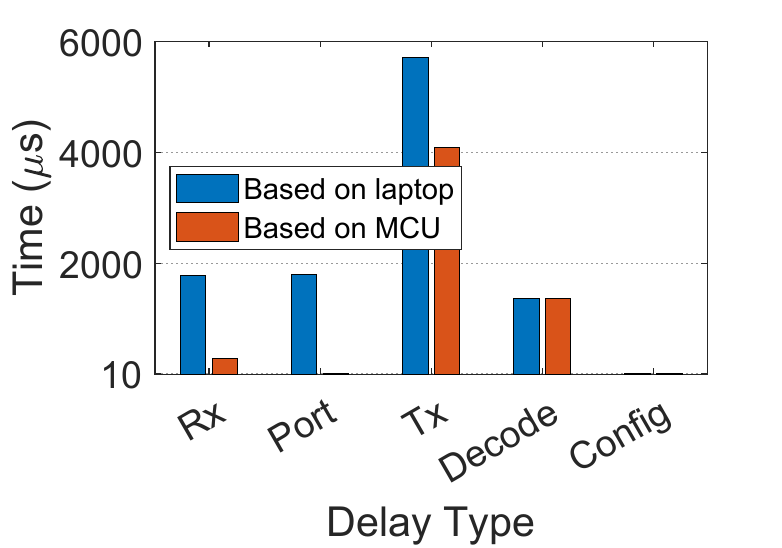}}\hspace{2mm}
	\subfloat[Comparison of ChannelDance, RBLE hopping delay and commodity BLE hopping interval.]{\includegraphics[width=0.46\linewidth]{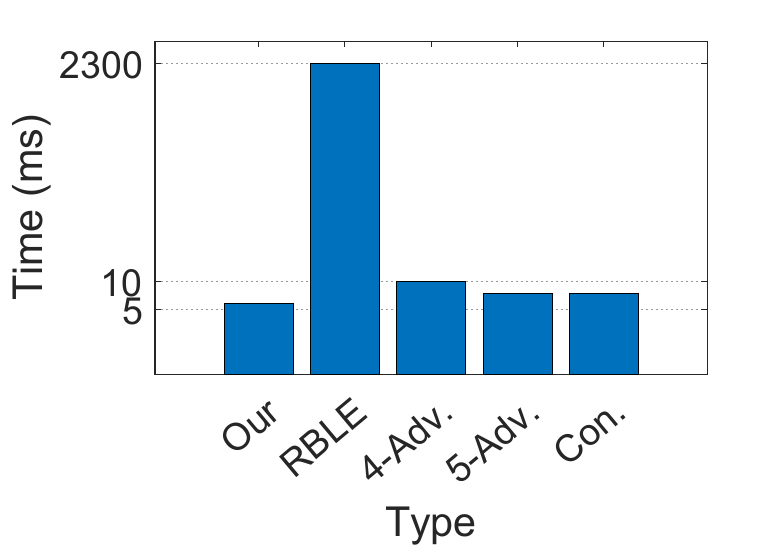}}
	\caption{Hopping delay comparison and analysis.}
	\label{delay}
\end{figure}

\textbf{Hopping delay}. ChannelDance edge server assists tag in doing adaptive channel hopping. We analyze and measure the delay of different parts. Meanwhile, we also compare ChannelDance hopping delay with RBLE and some common hopping intervals in commodity BLE. We implement the edge server with two architectures, laptop-based and MCU-based.  We let the edge server forward 20 Bytes data to the tag. As shown in Fig. \ref{delay}(a), in the Rx and Tx part, MCU-based architecture has an improvement of 1700 $\mu s$, due to the high-speed UART and SPI ports. Meanwhile, MCU-based have reduced the port forwarding time from 1800 $\mu s$ to 13 $\mu s$. Because MCU is more suitable for dedicated data processing. Since decoding the downlink signal is on the tag, both architectures consume the same time of 1300 $\mu s$. And it takes the tag 13 $\mu s$ to reconfigure its clock. The total time is about 5.8 $ms$.

Fig. \ref{delay}(b) shows the comparison between ChannelDance and other systems. RBLE adopts packet length modulation (PLM) on the downlink. The packet interval of the CC2540 which RBLE uses is about 14 $ms$, and the PLM symbol length on average is 240 $\mu s$. Thus it takes RBLE about 2300 $ms$ to transmit 20 Bytes data on the downlink. In commodity BLE, devices can advertise directed packets on the three advertising channels, with a hopping interval of less than 10 $ms$. And in BLE 5, the periodic advertisement hops on data channels and the minimum interval can be 7.5 $ms$. For BLE data connection, the minimum hopping interval of BLE 4 and 5 are all 7.5 $ms$. So ChannelDance’s low hopping delay meets the hopping requirements of most applications in BLE, and is 400x lower than RBLE's hopping delay.

\textbf{Resources consumption}.
We measure the consumed resources on the tag under a different number of precomputed clock states, which supports hopping. As shown in Table \ref{luts}, the number of LUT consumed increases as states increase. 
When ChannelDance stores 64  precomputed states, the number of consumed LUTs is about 569, far less than the 17600 LUTs that ZYNQ has.

\begin{table}[]\centering
	\centering
	\caption{Resources consumption of clock states}
	\label{luts}
	\begin{tabular}{ccccccc}
		\hline
		Num of Clock States & 2  & 4   & 8   & 16  & 32  & 64  \\ \hline
		LUTs                & 78 & 105 & 155 & 219 & 347 & 569 \\ \hline
		ZYNQ-7010 LUTs      & \multicolumn{6}{c}{17600}        \\ \hline
	\end{tabular}
\end{table}

\subsection{Adaptive Hopping}

We conduct experiments to evaluate how well tag performing BLE channel hopping algorithm is. The excitation signal in this experiment is the periodic advertisement packets generated by the edge server. We conduct the two BLE hopping algorithms and transmit one packet on each channel. Then we observe how many packets are backscattered on these target channels in the first 1000 hops to infer the num of hops on each channel. The used target channels are 17, 18, …, 30, and 31. There will be a theoretical value for packet num in each used channel. We show the results on channels 18-22.

\begin{figure}
	\centering
	\begin{minipage}{0.49\linewidth}
		\subfloat[{Channel selection algorithm \#1.}]{\includegraphics[width=\linewidth]{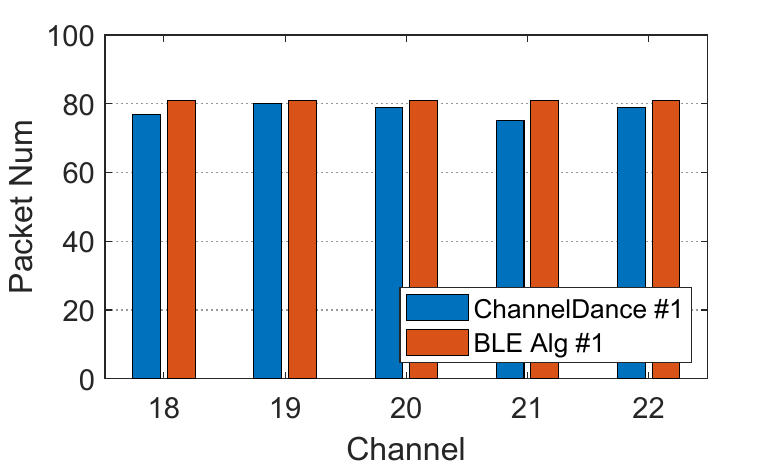}
			\label{fig-alg1}
		}
	\end{minipage}
	\begin{minipage}{0.49\linewidth}
		\subfloat[{Channel selection algorithm \#2.}]{\includegraphics[width=\linewidth]{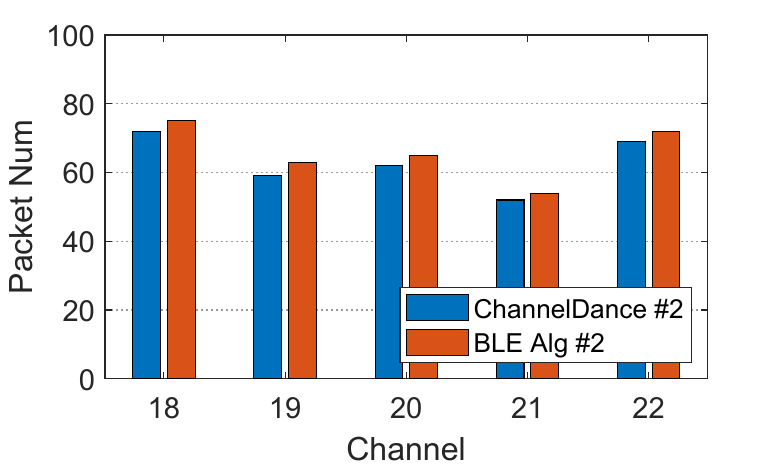}
			\label{fig-alg2}
		}
	\end{minipage}
	\caption{Comparison of ChannelDance hopping and commodity BLE.}
	\label{alg}
\end{figure}

\textbf{Channel selection algorithm \#1}. The channels are selected sequentially using a fixed hop interval. The expected packets on each channel are roughly the same. As shown in Fig. \ref{alg}(a), the expected packets on channels 18-21 are all 81 in the first 1000 hops. And our ChannelDance possesses the same pattern and achieves a success rate of at least 92\%.

\textbf{Channel selection algorithm \#2}. Algorithm 2 uses the current event counter (BLE hops across different events) and access address to calculate the target channel. The selected channel appears to be semi-random. As shown in Fig. \ref{alg}(b), the expected packet num varies on different channels. For example, the tag should hop to channel 18 for 75 times, and we observe that 72 packets are received. And the tag is expected to hop to channel 21 for 54 times and 52 packets are received. ChannelDance achieves a success rate of more than 93.6\%. Results show that BLE hopping algorithm \#2 is well performed on ChannelDance too.

\subsection{Application}
\subsubsection{Channel optimization}
\begin{figure}[t]
	\centering
	\subfloat[PER comparison of different target channels.]{\includegraphics[width=0.48\linewidth]{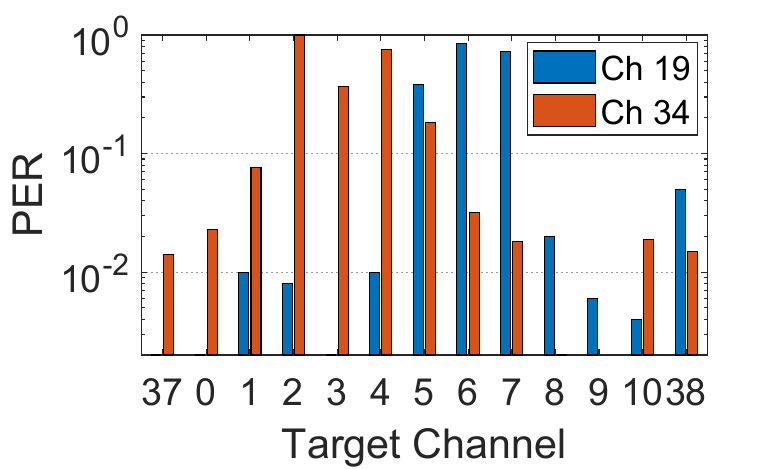}}\hspace{2mm}
	\subfloat[Goodput distribution with and without channel optimization.]{\includegraphics[width=0.48\linewidth]{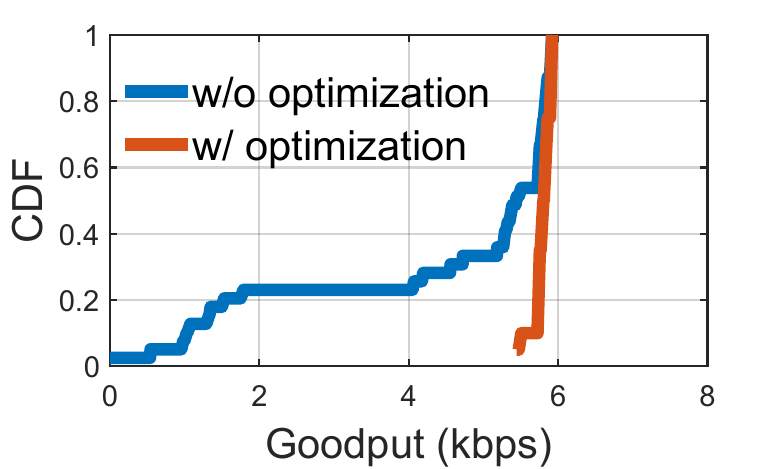}}
	\caption{Performance of channel scanning and optimization.}
	\label{PER}
\end{figure}

Different target channels could have different channel quality. We further measure the PERs during hopping.  As shown in Fig. \ref{PER}(a), we take two cases of excitation channels and show the PER performance among some target channels. When the excitation signal is on channel 19, the PER varies among different target channels. Channels 37, 7, 8, and 9 have small PERs, and channels 2, 4, and 5 have high PERs. While for excitation channel 34, the situation changes. Channels 37, 0, 1, 2, 3, 4, 10 have PERs less than 1\%. But Channels 2, 3, 4, and 5 suffer from a PER of larger than 10\%. 

With hopping, ChannelDance can exclude the “bad” channels. Fig. \ref{PER}(b) shows the goodput improvement on excitation channel 19. Before optimization, the goodput on 20 percent of the target channels is within 1.6 kbps under a packet interval of 50 $ms$. ChannelDance scan the 39 channels, calculate the median PER, and then exclude the target channels of which PER is less than the median. After utilization, 20 percent of the target channels are within 5.7 kbps, which achieves a 3.56x gain.
\subsubsection{BLE connection}
\begin{figure}[t]
	\centerline{\includegraphics[width=0.87\linewidth]{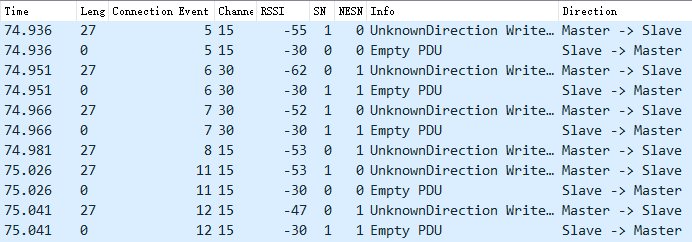}}
	\caption{Sniffer output of connection between ChannelDance tag and ambient BLE receiver.}
	\label{application}
\end{figure}

We build an application based on ChannelDance adaptive hopping. We make ChannelDance build a connection with an ambient BLE receiver, CC2540. The CC2540 advertises on advertising channel 38. ChannelDance edge server forwards the advertisement information to the tag. Meanwhile, it also provides periodic excitations. ChannelDance tag uses the advertising information to generate corresponding request packet and establish a connection with the receiver following BLE specifications. During connection, the involved devices exchange data on at least two hopping channels. The ChannelDance tag gets the excitation information from the edge server and meanwhile calculates the target channel based on the hopping algorithm. Thus it can choose different clocks to achieve adaptive hopping on the used channels 15 and 30 in different connection events, as shown in Fig. \ref{application}, and send WRITE packets. 

\section{Discussion and Future work}
\textbf{Multiple tags.} Multiple tags can simultaneously receive the downlink data to get the excitation channel information. But the tags can not simultaneously backscatter on the same target channel as they will interfer with each other. So each tag should be assigned an ID and the edge controls the tags to backscatter on different target channels. We will investigate a detailed MAC layer design for multiple tags in the future.

\textbf{BLE 5 compatibility.} We will explore more BLE 5 features including 2M PHY, extended advertisement and BLE mesh, which makes tags more compatible with BLE network.

\section{Related work}
Our system proposes a fast and accurate hopping design for Bluetooth backscatter, which is inspired by the recent progress on backscatter systems.

\noindent{\bf Backscatter communication.} 
Ambient backscatter\cite{liu2013ambient} leverages the TV signal as excitation signal, and it opens the door for backscatter transmission \cite{zhang2020reliable,zhang2017freerider,zhang2016hitchhike,zhang2017freerider,li2018passive,peng2018plora,talla2017lora,wang2017fm} using ambient signal as a carrier. 
WiFi Backscatter \cite{kellogg2014wi} enables the first general-purpose WiFi backscatter.
FS backscatter \cite{zhang2016enabling} observes that frequency shift is the key to improving the communication quality of backscatter transmission.
BackFi \cite{bharadia2015backfi} has improved the throughput to data rates of 5-300 Mbps. Passive WiFi uses a dedicated plug-in device to transmit single tones and achieves up to 11 Mbps data rate.
Hitchhike\cite{zhang2016hitchhike} firstly enables productive backscatter with commodity devices. It introduces the codeword translation technique to achieve.Interscatter\cite{iyer2016inter}provides a reverse-whitening technique to generate single tones using a Bluetooth device. 

\noindent{\bf BLE backscatter.}
BLEbackscatter\cite{ensworth2017ble} uses a CW generator to provide carriers for the BLE backscatter. FreeRider\cite{zhang2017freerider} extends codeword translation technique from WiFi to BLE. However, it suffers from an unreliable two-step modulation. Meanwhile, the backscattered packets cannot pass the CRC in the commodity receiver. To enable direct communication with a commodity BLE device, RBLE\cite{zhang2021commodity} presents a BLE backscatter system communicating with a single receiver.
To better improve the modulation performance, IBLE\cite{zhang2021commodity} introduces GFSK to improve modulation quality.

In summary, the above systems are working on the progress of turning BLE backscatter into a general-purpose communication system. In this way, we build ChannelDance to further support the channel hopping mechanism in BLE.

\section{Conclusion}
We present ChannelDance, a BLE backscatter system that utilizes a low latency edge server to achieve fast and accurate hopping. We propose an accurate hopping design on tag with a dynamically configured clock and identify channels of good quality. Further, we introduce an edge with a centralized and low latency architecture, which supports fast hopping of tag when communicating with commodity device. We build prototypes and conduct experiments to verify the feasibility and performance of ChannelDance. 
We believe that supporting accurate and fast hopping is significant to developing general-purpose communication for IoT since it 
can be incorporated into the existing BLE network.

\bibliographystyle{plain}
\bibliography{reference}

\end{document}